\documentclass[12pt,preprint]{aastex}

\usepackage{color}

\def\BB{{\bf {B}}}
\def\JJ{{\bf {J}}}
\def\vv{{\bf {v}}}
\def\xx{{\bf {x}}}

\shorttitle{Compression of potential magnetic fields}
\shortauthors{Pontin \& Huang}

\begin{document}

\title{On the formation of current sheets in response to the compression or expansion of a potential magnetic field}


\author{D.~I.~Pontin}
\affil{Division of Mathematics, University of Dundee, Nethergate, Dundee, DD1 4HN, UK}
\email{dpontin@maths.dundee.ac.uk}

\author{Y.-M.~Huang}
\affil{Space Science Center, University of New Hampshire, Durham, NH 03824, USA}

\begin{abstract}
The compression or expansion of a magnetic field that is initially potential is considered. It was recently suggested by Janse \& Low [2009, {\it ApJ}, {\bf 690}, 1089] that, following the volumetric deformation, the relevant lowest energy state for the magnetic field is another potential magnetic field that in general contains tangential discontinuities (current sheets). Here we examine this scenario directly using a numerical relaxation method that exactly preserves the topology of the magnetic field. It is found that of the magnetic fields discussed by Janse \& Low, only those containing magnetic null points develop current singularities during an ideal relaxation, while the magnetic fields without null points relax toward smooth force-free equilibria with finite non-zero current.
\end{abstract}



\section{Introduction}
Force-free magnetic fields are of great importance to our understanding of many astrophysical plasmas. In the solar corona, for example, where magnetic energy in general dominates over the kinetic, thermal, and gravitational energy, the magnetic field is typically considered to be approximately force free. A force-free magnetic field $\BB$ satisfies
\begin{equation}
\nabla\times\BB = \alpha\BB
\end{equation}
where $\nabla\times\BB/\mu_0=\JJ$ is the current density. In general $\alpha=\alpha(\xx)$, but if $\alpha$ is either spatially uniform or zero we distinguish linear-force-free fields or potential fields, respectively.
Two magnetic fields on the same domain are said to be {\it topologically equivalent} if one can be obtained from the other by means of a smooth (continuously differentiable) deformation.
The topology of a magnetic field in a magnetically closed volume is therefore specified by the orientation and mutual linkage of all field lines. This encompasses the presence of field nulls (points at which $\BB={\bf 0}$) and separator fields lines linking such nulls. During an ideal evolution of the magnetic field (which includes the evolution of $\BB$ in an ideal plasma), this topology is preserved \citep[for a detailed discussion see][]{hornig1996}. When the volume under consideration is not magnetically closed -- i.e.~when field lines thread the boundaries of the volume -- then one must refer to the topology with respect to some external reference field. In this case, conservation of the topology requires that the mapping of field lines from one boundary of the domain to another is fixed or follows some ideal evolution of the external field -- in addition to the preservation of field line linkage and numbers of null points.

Explaining how the solar corona is heated to temperatures of millions of degrees Kelvin above that of the solar surface is a long-standing and fundamental problem in solar physics.
One leading, though still controversial, mechanism  is the ``topological dissipation" model put forward by \cite{parker1972}. This model is based on the assertion that if one chooses a magnetic field of a given topology, then in most cases no smooth force-free equilibrium corresponding to that topology exists. Therefore the lowest energy state will in general be a force-free field containing tangential discontinuities, i.e.~current sheets.
Since the Alfv{\' e}n transit time along a loop is much shorter than timescales associated with the convective motions in the photosphere, it is argued that the coronal field evolves through a sequence of (approximate) force-free states. As the boundary driving continually changes the topology of the field, one will inevitably arrive at a topology for which no smooth force-free state exists, leading to the formation of current sheets. Of course in the real coronal plasma the small but finite resistivity means that the truly discontinuous state is never reached, but once length scales are sufficiently small the current will be dissipated, accompanied by magnetic reconnection and a change of topology.

One reason that the topological dissipation mechanism has stimulated so much debate is that so far no-one has been able to construct a general proof for or against it. Many studies have tackled different aspects of the problem using both theoretical and numerical techniques \citep[e.g.][]{vanballegooijen1985,zweibel1987,longcope1994,parker1994,ng1998,craigsneyd2005}.
A series of recent studies have attempted to place the hypothesis on a firmer theoretical footing by considering ``topologically untwisted" force-free fields  \citep{low2006,low2007,janse2009} -- which according to the authors' definition equate with potential fields in the case of a closed, bounded, simply-connected domain. In particular, \cite{janse2009} (hereafter referred to as JL09) considered the compression or expansion of a magnetic field that is initially potential and is line-tied between two perfectly conducting plates. The normal component of the field is taken to be non-zero only on these two line-tied plates, and the system is perturbed by moving the two line-tied boundaries either closer to or further from one another. JL09 argue that since the field is topologically untwisted, and since no twist has been applied via the boundary perturbation, then the field must remain topologically untwisted, i.e.~it must remain potential. They go on to present specific examples where they calculate the new potential field in the compressed (or expanded) volume and demonstrate that its topology is not the same as that of the initial field. Thus, they argue, if a relaxation ensues in which the topology is preserved, current sheets must form. 
We note that this type of finite-amplitude compression is rather different from the original Parker mechanism, which invoked only random, small-amplitude footpoint motions tangential to the line-tied boundaries. In the study of JL09, the choice of a compression of the volume was simply a method envisaged by the authors to generate a domain and magnetic topology that are inconsistent with the existence of a smooth equilibrium.
It is worth noting that the formation of current singularities in response to the compression of a force-free field was also considered by \cite{bobrova1979}. However, they considered a scenario in which the magnetic field was one-dimensional, and of infinite extent, being tangential to the perfectly conducting boundaries.

There have been two criticisms made of the JL09 argument. First, \cite{huang2009} presented a specific counter-example in which the magnetic field, once compressed, did not lead to a potential field with current sheets, but rather to a smooth force-free field with non-zero current. This they demonstrated in the linear regime by using an expansion scheme, and also numerically in the non-linear regime. However, \cite{janse2010} have pointed out that this example does not contradict their theory, since the magnetic field considered by \cite{huang2009} is two-dimensional, being unbounded in the invariant direction.

\cite{aly2010} have also raised an objection against the argument of JL09. They consider the same geometry as JL09, and assert that one step in the JL09 argument is flawed. In particular, they draw attention to the fact that the assumption that the compressed magnetic field must remain topologically untwisted is based on an intuitive rather than a firm theoretical basis. Indeed, they argue that the field must not remain potential (topologically untwisted), but must develop large-scale finite currents. 

In this study, we use a direct approach to investigate the question as to whether current singularities form during an ideal relaxation. We take a given potential magnetic field over a closed 3D domain as initial condition, deform the domain by compressing/expanding it by some chosen factor, and then numerically relax the magnetic field towards a state with $\JJ\times\BB={\bf 0}$. We then analyse the properties of the magnetic field and current density obtained in the final state of the relaxation. In one particular example, we take as an initial condition the magnetic field without null points discussed by JL09. We describe the numerical method in Section \ref{method}, then present our results in Section \ref{results} and a discussion in Section \ref{discuss}.

\section{Numerical method and signatures of singularity}\label{method}
We use a 3D Lagrangian relaxation scheme \citep[see][]{craig1986}, which takes  advantage of the fact that, 
due to the frozen-in field condition, the evolution of the vector field ${\bf B} / \rho$  is described by exactly the same
equation as the evolution of the line element $\delta {\bf x}$ joining two Lagrangian
fluid particles ($\BB$ is the magnetic field and $\rho$ the mass density).  Specifically,
\begin{equation}
\frac{D}{Dt} \left(\frac{{\bf B}}{\rho}\right)  = \left( \frac{{\bf
B}}{\rho} \cdot \nabla \right) {\bf v},
\end{equation}
where $D/Dt$ denotes the material derivative and $\vv$ the plasma velocity.  Since we are only interested in the final equilibrium configuration, and not in the evolution to it, we
determine the relaxed state using a fictitious damping force. Specifically, the evolution of the locations (${\bf X}$) of the fluid particles -- which are used as grid points in our Lagrangian approach --  is determined by 
\begin{eqnarray}
\frac{D {\bf X}} {Dt} & = & {\bf J} \times {\bf B} 
\label{momentum}
\end{eqnarray} 
where  ${\bf J}= (\nabla \times {\bf B}) $ is the dimensionless
current density. This is equivalent to neglecting inertial effects and introducing a frictional force into the momentum equation, where the coefficient of friction is set to 1. Such a choice has the advantage of eliminating the wave properties of the plasma while guaranteeing a monotonic decay
of the total energy. Moreover, it can be shown that the path taken in
the relaxation affects neither the final equilibrium attained nor its
stability properties \citep{chodura1981, craig1986}.

During the relaxation process, fluid elements are evolved according to Equation (\ref{momentum}), while we fix the locations of the fluid elements on all boundaries to impose line-tying of the magnetic field. The  magnetic
flux, magnetic helicity, and $\nabla \cdot {\bf B}$ are automatically
conserved in the relaxation.  Thus, if the initial magnetic field
satisfies $\nabla \cdot {\bf B}=0$, this will always remain satisfied.
The numerical scheme is implicit and unconditionally stable, meaning  that the
computational cost of the simulations need not be prohibitive.

The purpose of this study is to search for the formation of tangential discontinuities -- current sheets -- in the relaxed field. 
While such current sheets cannot be explicitly represented on a discrete numerical grid,
there are certain signatures that are indicative of the presence of singular structures in the magnetic field represented by the discretisation. 
One indication of a potentially singular current layer would be the presence of current structures with thickness down to the grid scale.
In addition, we expect that  local quantities such as
the current density may diverge with increasing numerical resolution at certain points in the domain.
By contrast,  the integrated current  should remain  constant as the resolution is varied,   since this may be determined by
a line integral (via Amp{\` e}re's law) over a circuit far from the current buildup.
On the basis of a one-dimensional analysis it is
expected that the divergence of the current density with increasing resolution should take the form
\begin{equation}\label{scale}
J_{max} \simeq  J_0 N^{\mu}
\end{equation}
where $\mu>0$ is of order unity and $ N $ is the numerical resolution which defines the initial grid spacing according to $\Delta x \sim 1/N$ \citep{ali2001}. In particular, for a one-dimensional collapse with a Lagrangian mesh \cite{ali2001} showed that one expects a power-law exponent $\mu=2$. Such a power-law scaling has also been observed for more complex singular structures in 2D and 3D in previous studies employing a Lagrangian approach \citep{craiglitvinenko2005,pontincraig2005}, for which typically $0<\mu<2$.

\section{Numerical relaxation results}\label{results}
\subsection{Magnetic fields without null points}\label{nonullsec}

\subsubsection{Example 1: The magnetic field of \cite{janse2009}}\label{janseexsec}

We first consider the three-dimensional potential magnetic field without null points discussed by JL09 (see Section 3.5.1 of their paper). This magnetic field is defined in cylindrical polar coordinates $(r,\theta,z)$ by
\begin{equation}\label{nonulleq}
\begin{array}{ccl}
\BB &=& \nabla\Phi\\
\Phi &=& B_0 z + J_1(\kappa r) \left(
\displaystyle\frac{\cosh(\kappa z)}{\kappa \sinh(\kappa L)}\sin\theta + \frac{\sinh(\kappa z)}{\kappa \cosh(\kappa L)}\cos\theta
\right),
\end{array}
\end{equation}
where $\kappa=k_{1,1}\approx 1.8412$ is the first root of the derivative of the 1st-order Bessel function $J_1$, and $B_0=1.3$. The authors consider this field over the domain $-L\leq z\leq L$, $r<1$. All field lines enter the volume at $z=-L$ and exit at $z=L$, and are assumed to be line-tied at these two planes. 

Since the numerical method we use is based on a Cartesian geometry, we consider this field, before any deformation, over the domain $-1.5 \leq x,y \leq 1.5$, $-1\leq z\leq 1$. We focus in what follows on the behaviour within $-1\leq x,y\leq 1$. It has been checked that the influence of the side boundaries does not affect the behaviour of the system within this region, as discussed below. We then compress/expand the domain in the $z$-direction in a linear fashion to disturb the magnetic field from equilibrium, before beginning the relaxation. Below we present results based on a compression by factor 1.25 such that the field is contained within $-0.8 \leq z \leq 0.8$. However, we have repeated the relaxation both for expansions, and for compressions up to a factor of 6.67 (such that $-0.15\leq z\leq 0.15$) and obtained qualitatively the same results. (For stronger compressions the relaxation scheme does not converge and so no conclusions can be drawn.)

At $t=0$ the peak value of the Lorentz force $|\JJ\times\BB|$ in the domain is around 1.15. We run the relaxation scheme as described above until this peak value falls below some threshold, here chosen to be $10^{-3}$ (at which point $|\BB|\approx 1.3$ and $|\JJ|\approx 0.035$). This value can be further decreased, but the qualitative results of the study remain the same.  We run the simulation for a series of numerical resolutions, where the number of grid points $N$ is the same in each coordinate direction. For a discussion of numerical errors see Appendix \ref{appA}, and for a further verification of the numerical method see the following section.

\begin{figure}[t]
\centering
(a)~~~\includegraphics[scale=0.7]{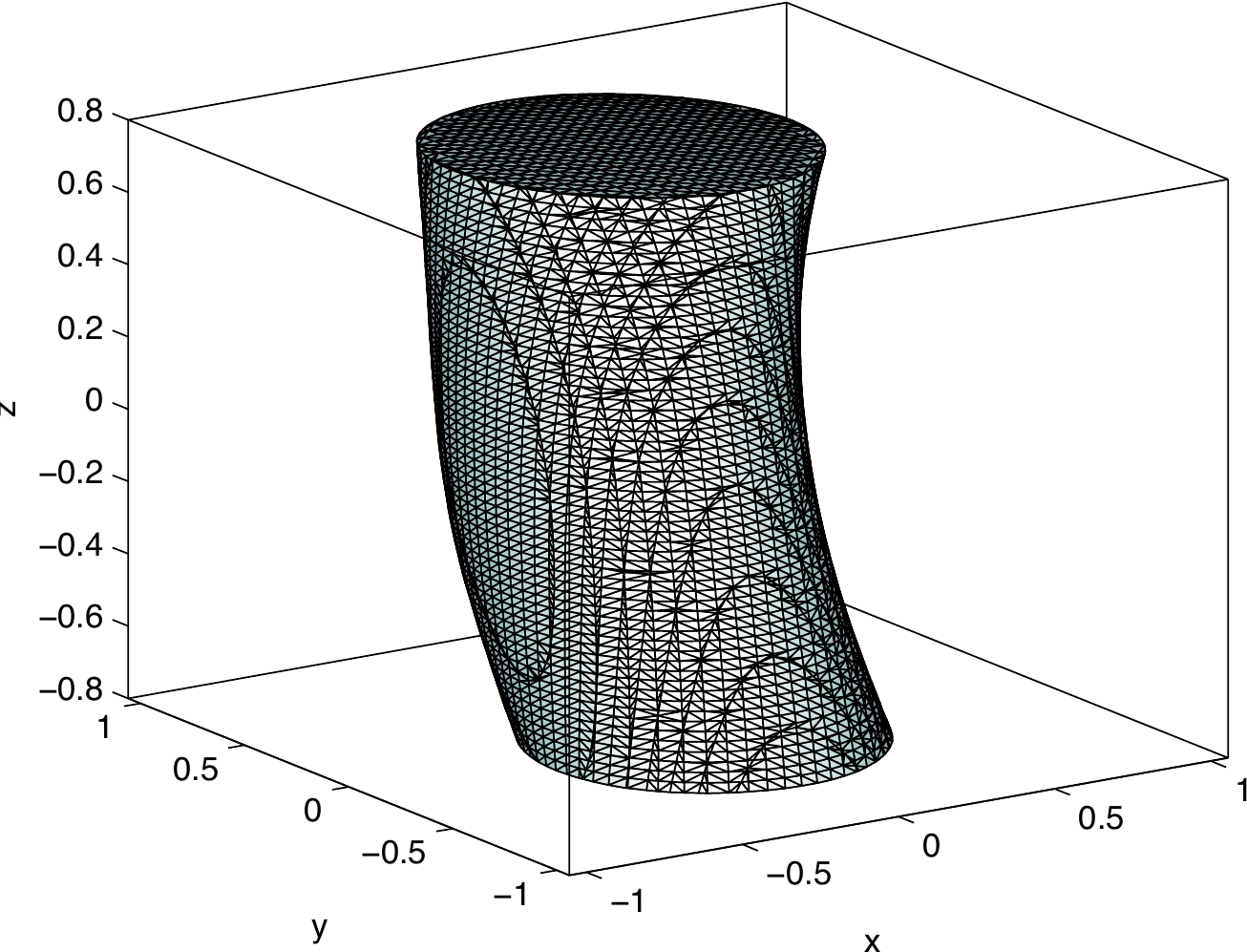}\\
(b)\includegraphics[scale=0.4]{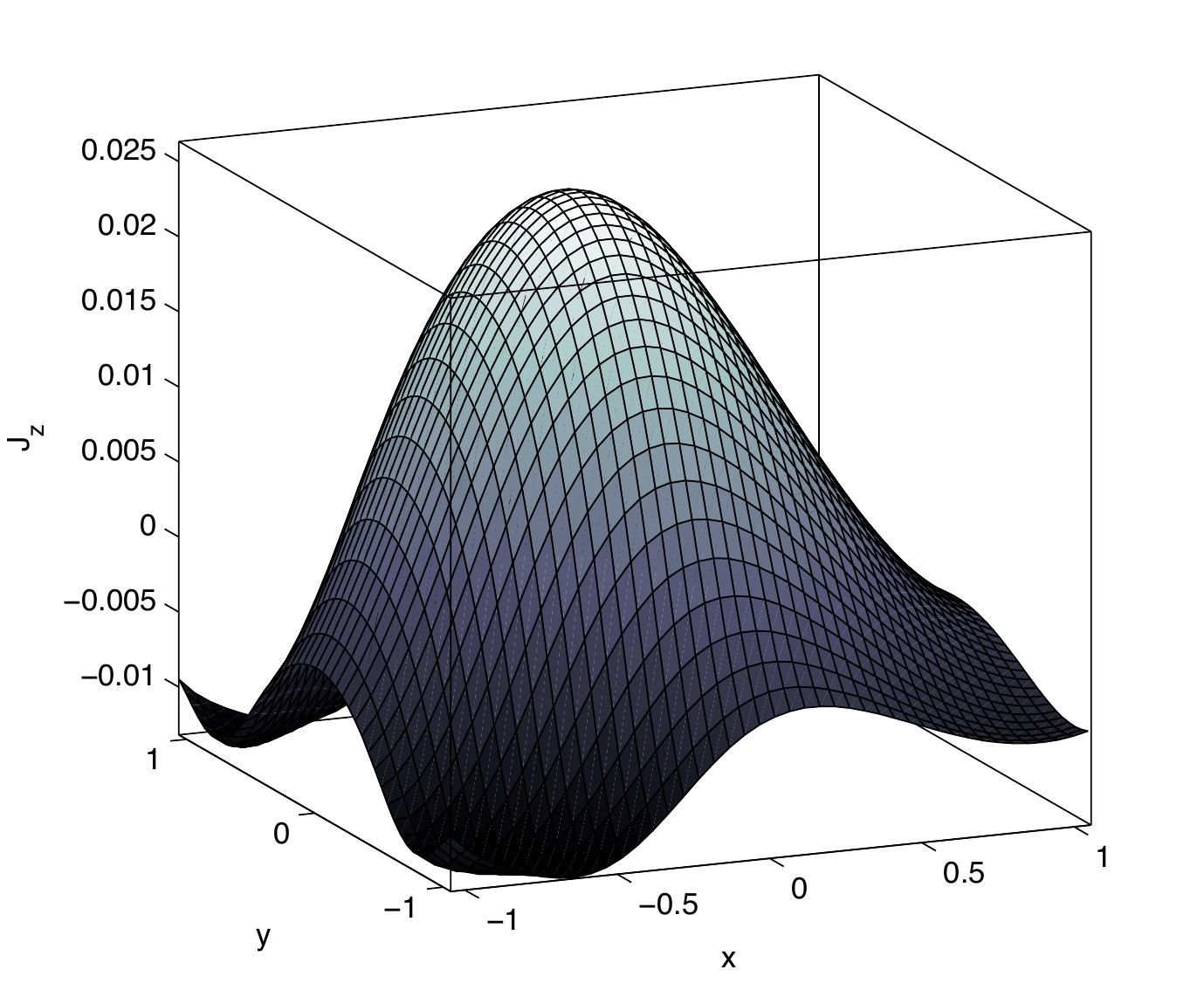}
(c)\includegraphics[scale=0.4]{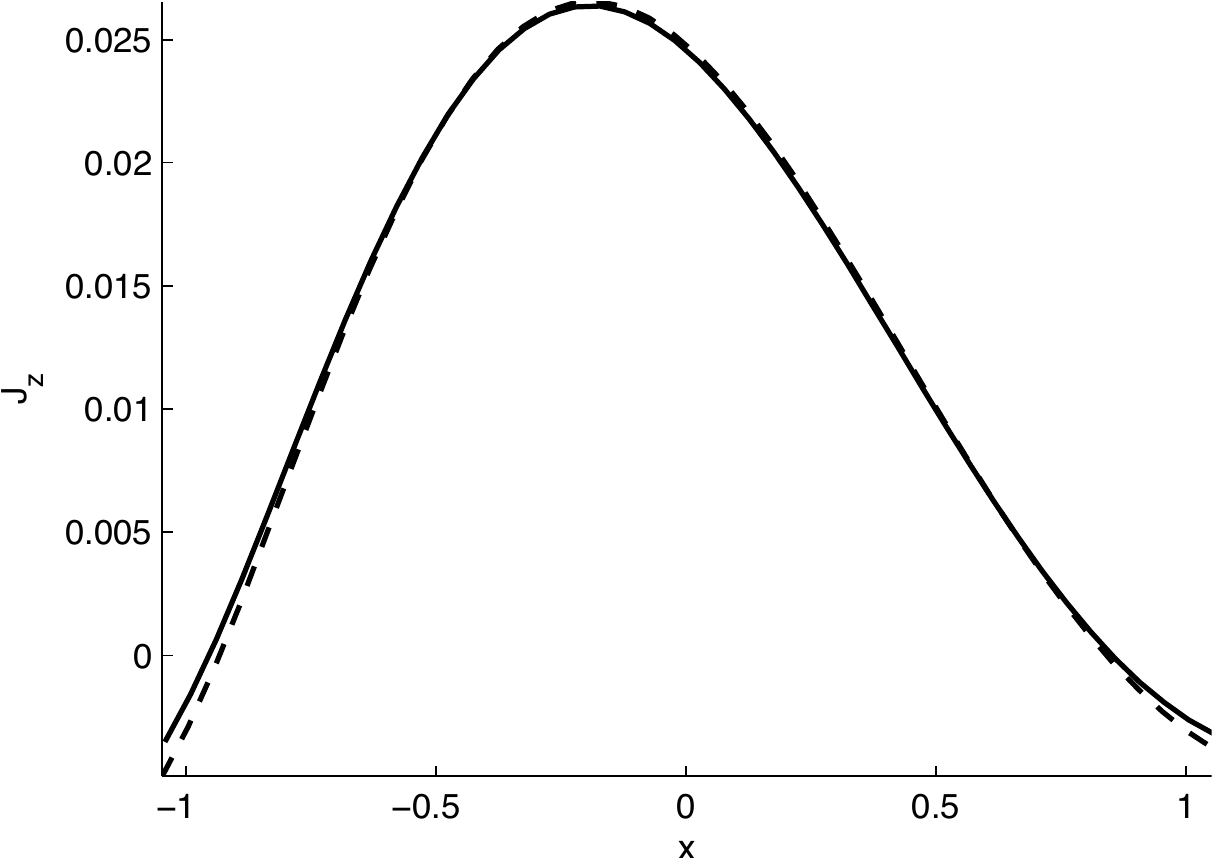}
\caption{(a) Current contour at 50\% of maximum for the relaxed state of the magnetic field without nulls (Eq.~(\ref{nonulleq})), from the simulation with $N=61$. (b) $J_z$ in the $z=0$ plane for the same simulation -- note that $J_z$ is significantly larger than the other components of $\JJ$. (c) $J_z$ along $y=z=0$ for the same simulation (solid line) and for a simulation with equivalent resolution over domain $-2\leq x,y\leq 2$ (dashed line).}
\label{jnn}
\end{figure}
\begin{figure}[t]
\centering
\includegraphics[scale=0.7]{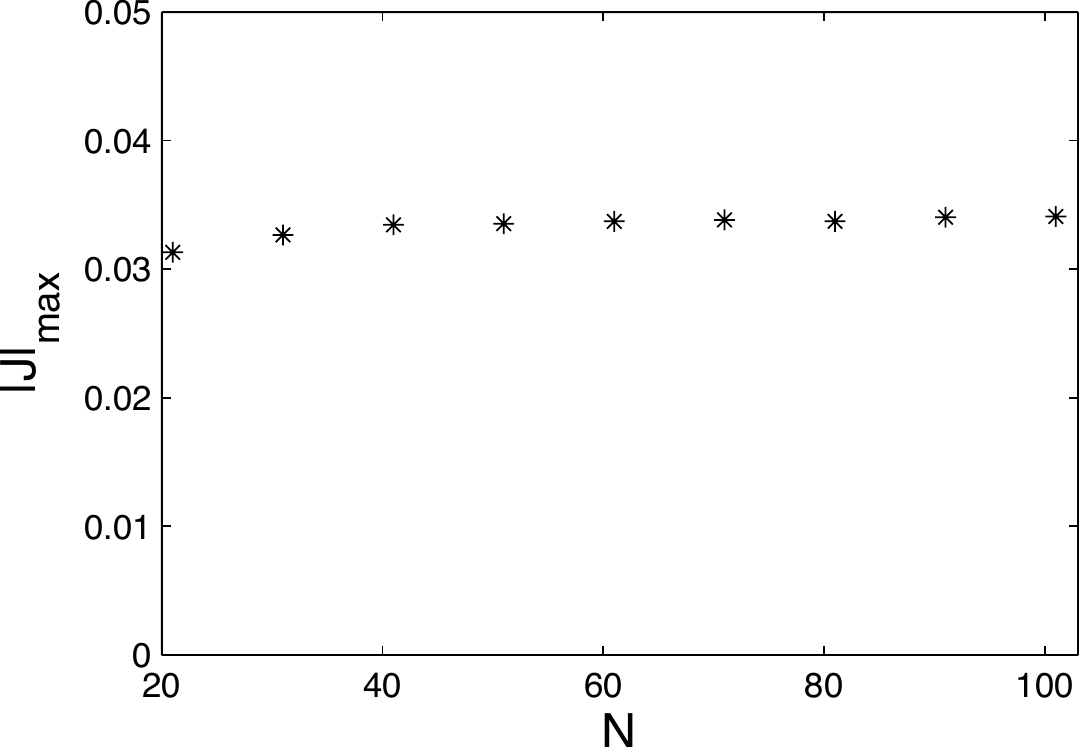}
\caption{Peak current density as a function of the numerical resolution $N$ for the relaxed state of the magnetic field without nulls (Eq.~(\ref{nonulleq})).}
\label{scalenn}
\end{figure}

In the relaxed state, a smooth equilibrium is obtained, with no indication of the formation of current sheets. The current density is non-zero over the vast majority of the volume, forming a large-scale tube-like structure extending over the length of the domain, as shown in Figure \ref{jnn}(a). The current density has a smooth profile within the domain (Figure \ref{jnn}(b)), and its properties are not significantly affected by the line-tied $x$- and $y$-boundaries, which can be seen by comparing the results of the simulation as run for different box sizes (Figure \ref{jnn}(c) -- this boundary condition being different to that considered by JL09, see above). Further evidence that no current sheets form during the relaxation is provided in Figure \ref{scalenn}, where it is demonstrated that the peak current in the relaxed state is essentially independent of the numerical resolution.

\subsubsection{Example 2: The magnetic field of \cite{huang2009}}

As an additional example, we have performed the same procedure as above for the magnetic field of \cite{huang2009} (which also contains no null points),
which was originally presented as a (two-dimensional) counter-example to the claim of JL09.  The initial magnetic field is
\begin{equation}\label{hbzb}
\BB=1\,{\hat {\bf x}}+{\hat {\bf x}}\times\nabla\left(y+\epsilon \sin(y) \cosh(z-1/2)\right)
\end{equation}
over $0\leq z \leq 1$. In the study of \cite{huang2009} the magnetic field is periodic in $y$ and assumed to have infinite extent in $x$. Here we enclose the domain within $-10\leq x,y\leq 10$ and $0\leq z\leq 1$ and discretise the domain with $121\times 121\times 41$ points. The $x$ and $y$ boundaries are sufficiently far away that they do not affect the solution within the domain of interest. We take $\epsilon=0.5$ and compress the field such that $0\leq z\leq 0.9$ before performing the relaxation. 
\begin{figure}[t]
\centering
\includegraphics[scale=0.5]{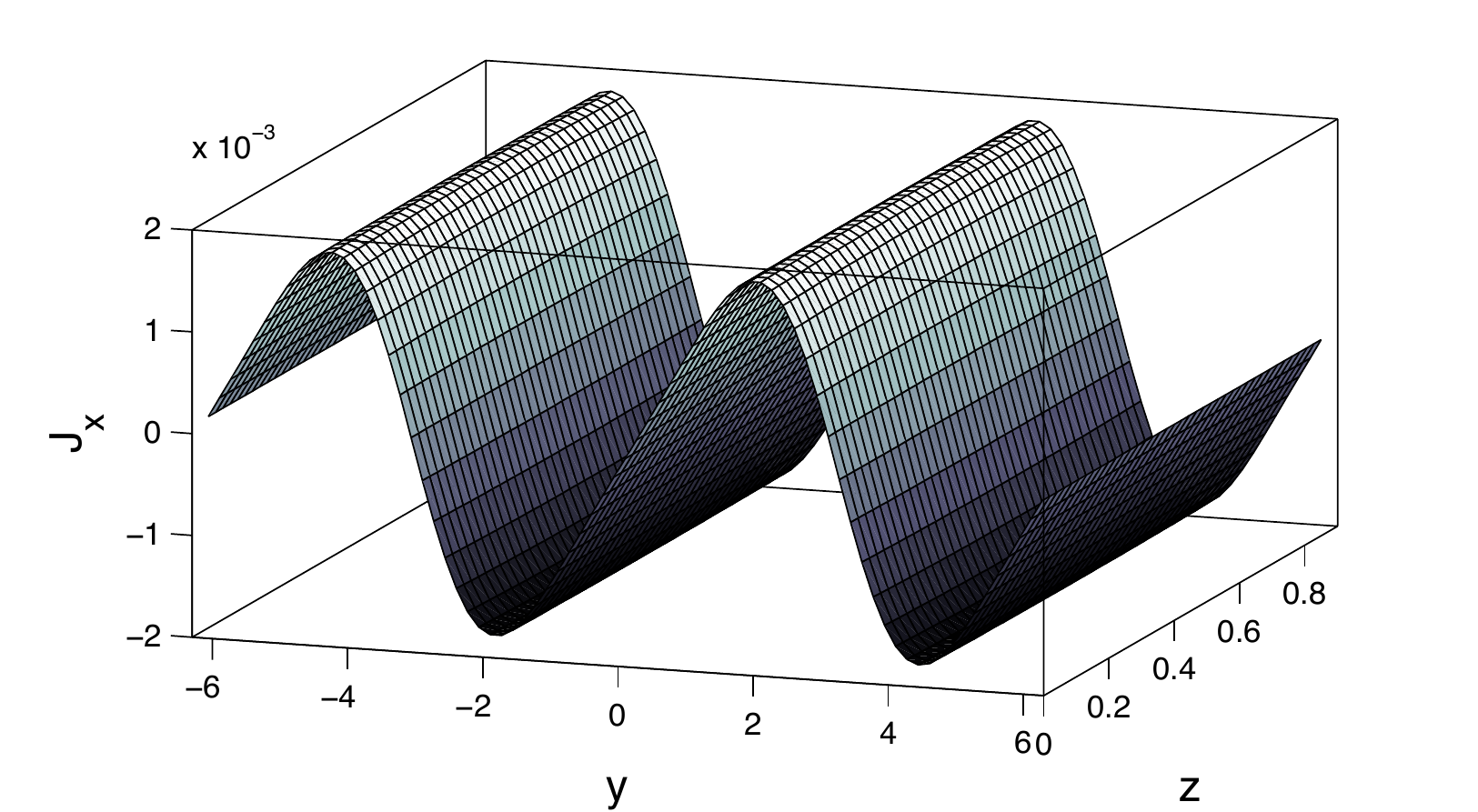}
\includegraphics[scale=0.5]{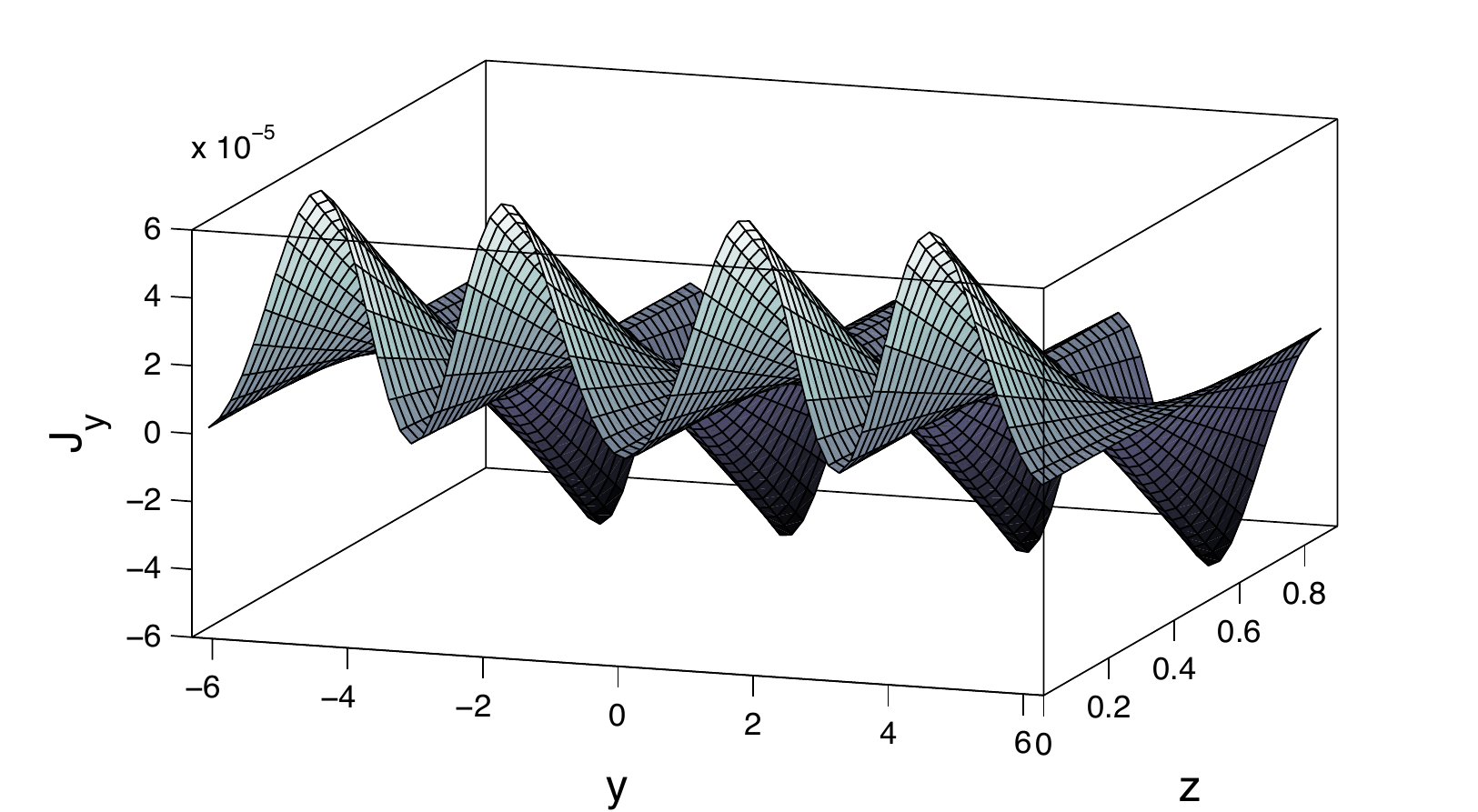}
\includegraphics[scale=0.5]{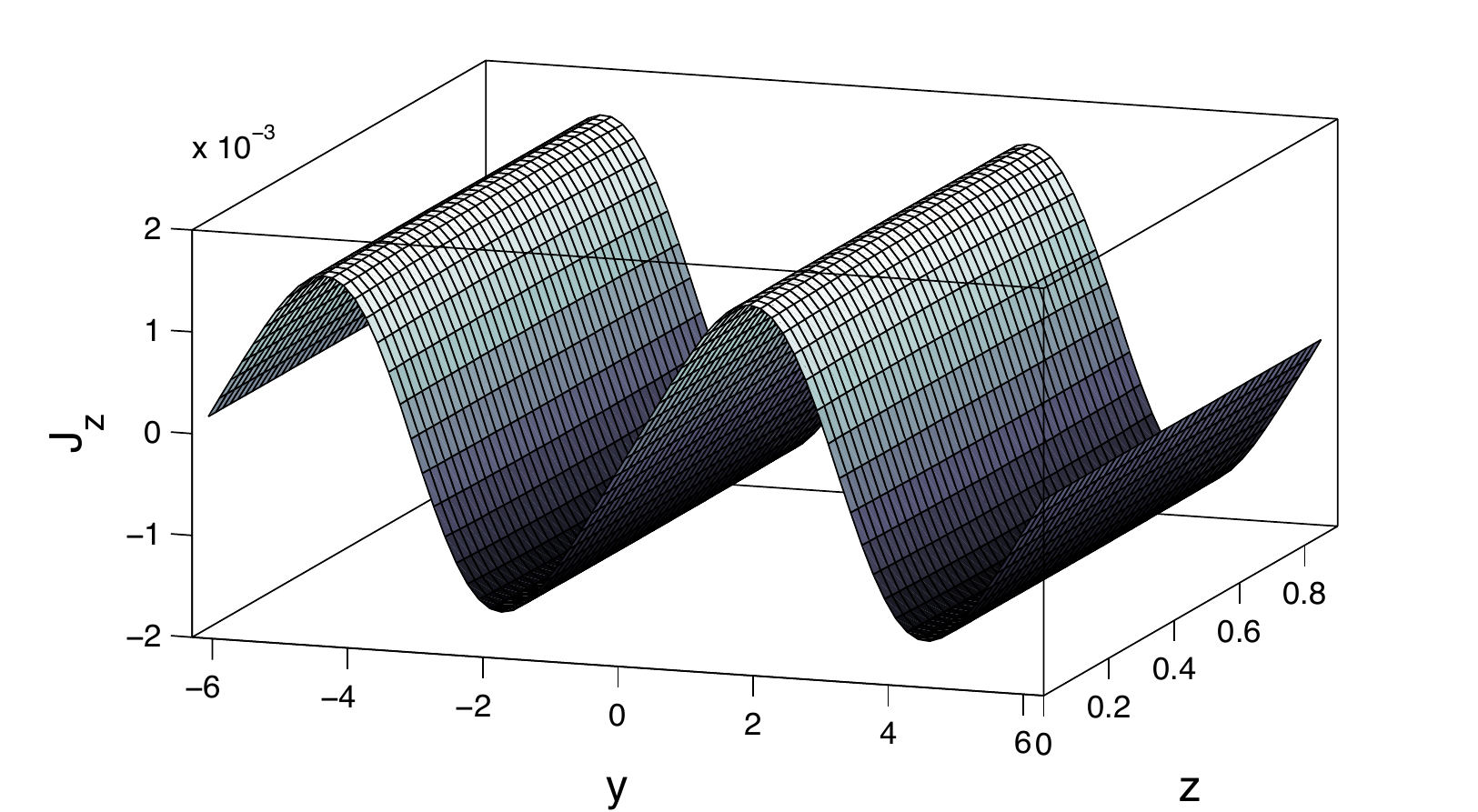}
\caption{The three components of the current density in the $x=0$ plane for the relaxed state of the magnetic field given in Eq.~(\ref{hbzb}).}
\label{hbzj}
\end{figure}

The results of the relaxation are qualitatively the same as in Example 1 presented above. A large-scale smooth current is distributed throughout the domain in the relaxed state, with no indication of small scales developing or the peak current locally diverging anywhere within the domain, as shown in Figure \ref{hbzj}. (Note that the largest currents within the domain are found in the volume containing field lines that penetrate the $x$-boundaries -- however they remain confined close the the boundaries for different simulation box sizes, and also show no signatures of singularity as discussed above.)
\begin{figure}[t]
\centering
\includegraphics[scale=0.6]{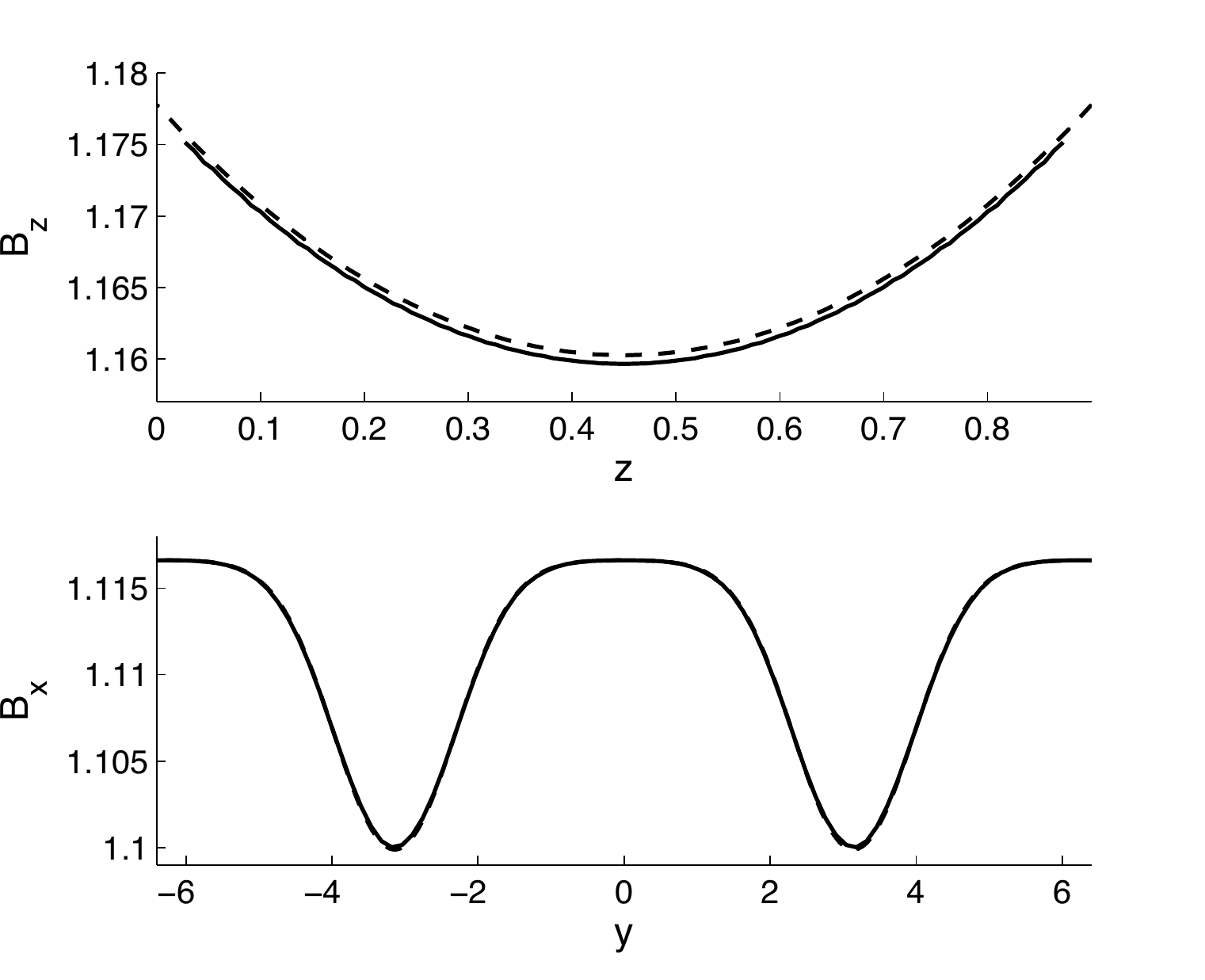}
\caption{Comparison between our numerical solution (solid lines) and the numerical solution presented by \cite{huang2009} (dashed lines) for the relaxed state of the magnetic field given by Eq.~(\ref{hbzb}). (a) $B_z$ along the line $x=0, y=-1.25$, and (b) $B_y$ along the line $x=0, z=0.4$. }
\label{hbzcomp}
\end{figure}
A comparison of our solution with the relaxed state obtained by \cite{huang2009} provides a further test of the accuracy of our relaxation method. The two numerical solutions are compared  along two representative straight line cuts through the volume in Figure \ref{hbzcomp}. The excellent agreement between the two methods is evident from the plots (indeed the two curves are almost indistinguishable for the plot of $B_x$ against $y$). The very small differences observed progressively decrease as we increase the grid resolution of our simulations.

\subsection{Magnetic fields containing null points}\label{nullsec}

\begin{figure}[t]
\centering
(a)\includegraphics[scale=0.5]{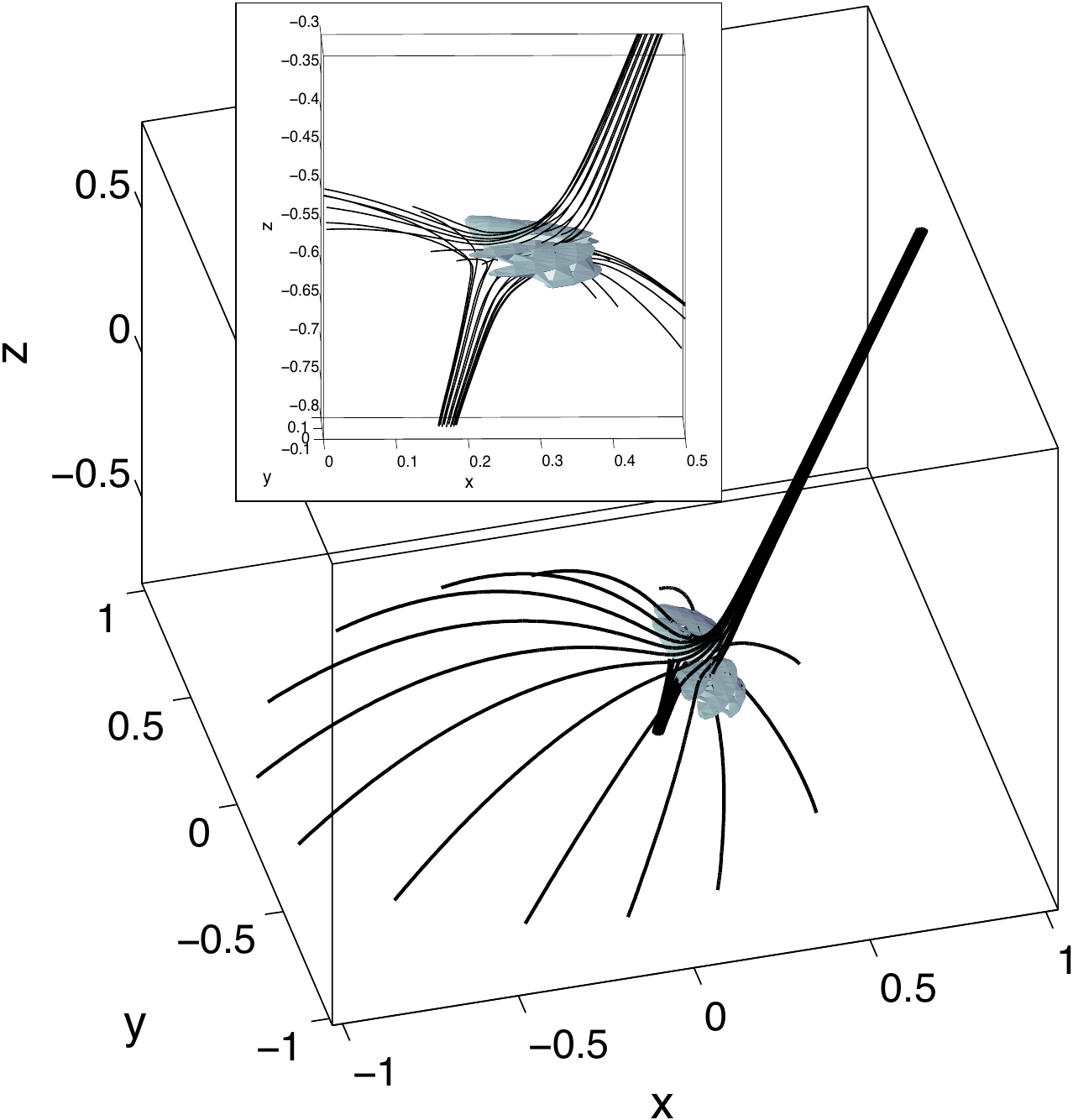}
(b)\includegraphics[scale=0.5]{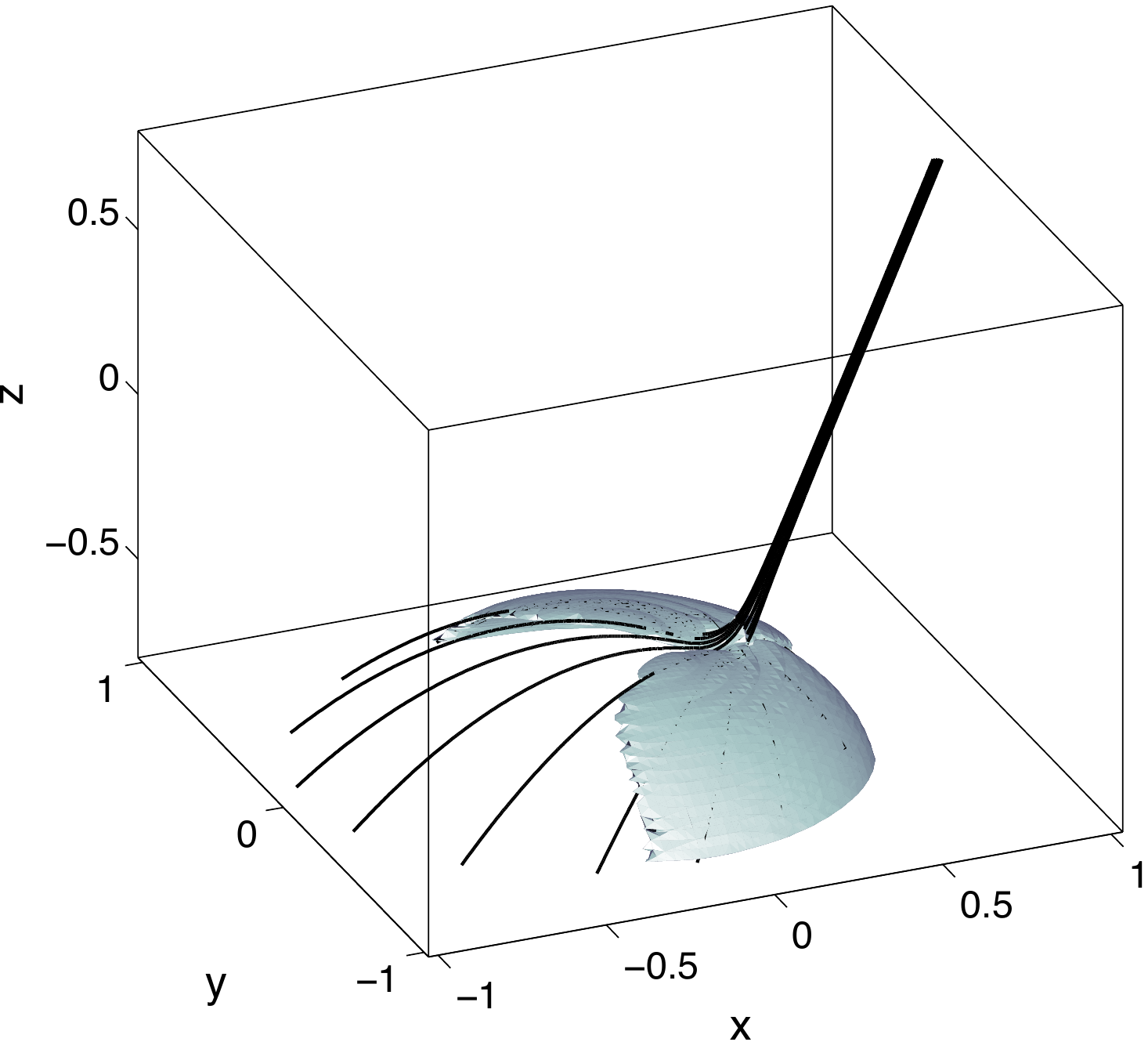}
\caption{Current contours at (a) 25\% of $|\JJ|_{max}$ and (b) $5\%$ of $|\JJ|_{max}$, for the relaxed state of the field containing a null point (Eq.~(\ref{nulleq})). Black lines are magnetic field lines, traced from close to the null. For the simulation with $N=81$.}
\label{jnull}
\end{figure}

We now consider an initial magnetic field that shares many properties with that described in Section \ref{janseexsec} above, but which contains a magnetic null point. We construct this potential field by superimposing onto a uniform background field a magnetic point source -- the point source is located outside the numerical domain so that the field within the domain is still physically plausible. Specifically, we take (in Cartesian coordinates)
\begin{equation}\label{nulleq}
\BB=[0.4,0,1]-\frac{0.6}{(x^2+(y-0.02)^2+(z+1.4)^2)^{3/2}}[x,y-0.02,z+1.4]
\end{equation}
over the domain $-1.5\leq x,y\leq 1.5$ and $-1\leq z\leq 1$. This magnetic field contains a null point located at $[x,y,z]=[0.277,0.02,-0.707]$. The fan separatrix surface of the null divides the volume into two distinct regions, and forms a dome-like structure (see Figure \ref{jnull}), beneath which the magnetic field lines are closed, i.e.~they both enter and exit the domain through the $z=-1$ plane. 
Note that we do not use the potential field with a null presented by JL09. Their field is more difficult to study with our numerical method since the field does not thread the top of the box, meaning that $|\BB|$ is much stronger near the base, and in addition the null is located very close to the base. We consider the magnetic field defined above to be more closely analogous to the fields without nulls presented in Section \ref{nonullsec}, in that the field threads both $z$-boundaries.

The magnetic field is perturbed, as above, by performing a linear compression of the numerical mesh such that after the compression $-0.8\leq z\leq 0.8$. This leads to a peak Lorentz force in the domain with value around 5. We then run the relaxation scheme until such time as $|\JJ\times\BB|<10^{-3}$ everywhere in the domain, and take the resulting field as the relaxed state.

\begin{figure}[t]
\centering
\includegraphics[scale=0.7]{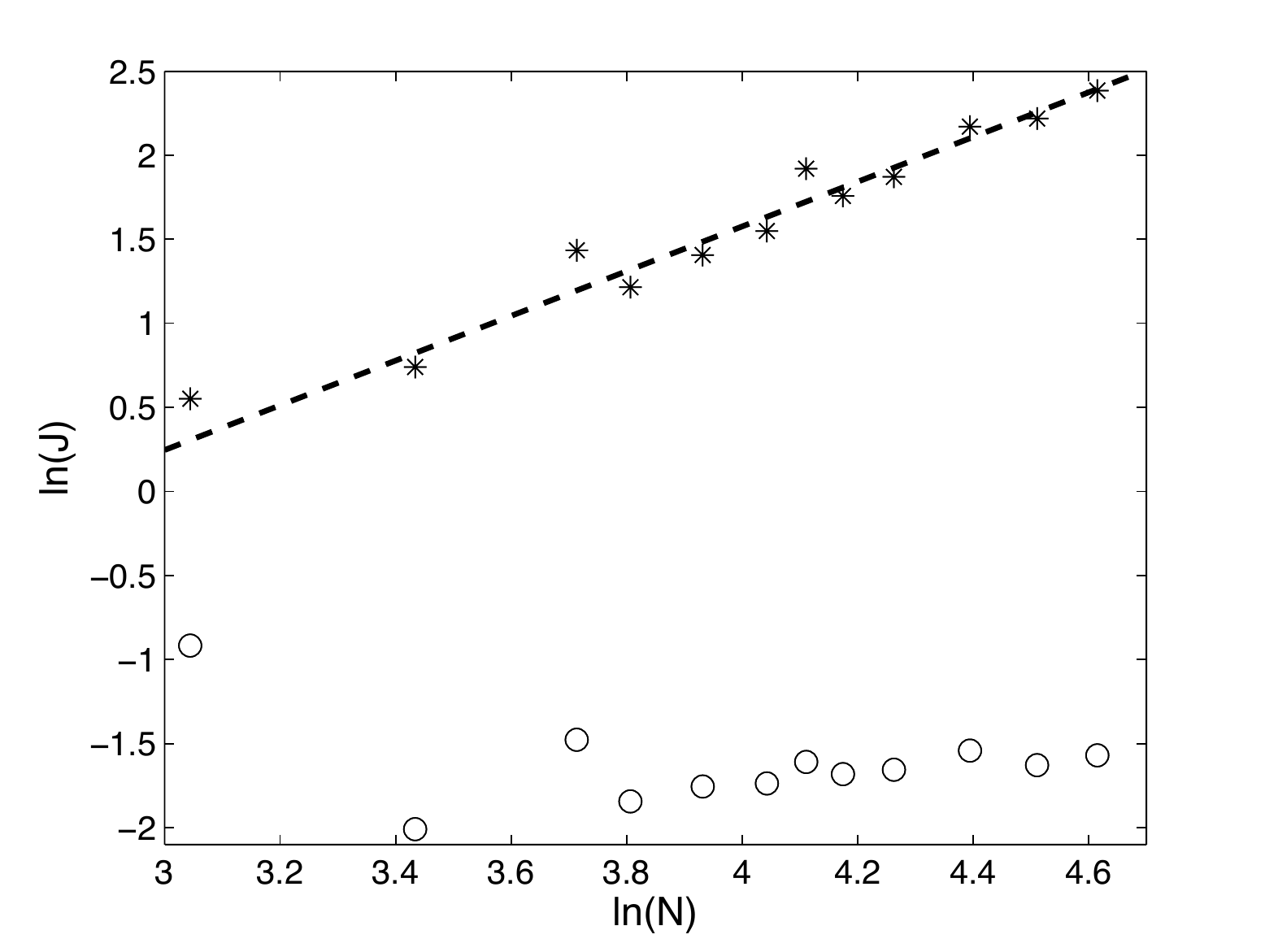}
\caption{Peak current density (stars) and integrated current density through the $y=0$ plane (circles) as a function of the numerical resolution $N$, for the relaxed state of the field containing a null point (Eq.~(\ref{nulleq})). The dashed line has a gradient of 1.33.}
\label{scalenull}
\end{figure}

In contrast to the field without nulls, there are clear signatures of current sheets in the relaxed state. First, a thin layer of intense current is found to form with thickness on the grid scale. This current layer is focussed at the null point, and spreads along the separatrix surface, as shown in Figure \ref{jnull}. This is consistent with the results of \cite{pontincraig2005} who investigated the perturbation of a simple linear null point magnetic field. Indeed, the local structure around the null is very similar (see the inset of Figure \ref{jnull}(a)), in that a local collapse of the spine and fan towards one another forms a current sheet centred on the null, with the current vector directed locally perpendicular to the plane of collapse \citep{pontincraig2005,pontinbhat2007a}. Furthermore, the current density within the sheet is found to diverge as the numerical resolution $N$ is increased. In Figure \ref{scalenull} we plot the peak current (found in the vicinity of the null) and the integrated current through the $y=0$ plane. It is clear that while the integrated current approaches a constant value, the peak current shows a power law dependence on $N$, consistent with the presence of an underlying singularity. The dashed line in Figure \ref{scalenull} indicates a scaling relation $J_{max}\propto N^{1.33}$, which is in line with the results of \cite{pontincraig2005} who considered only a linear null point field. We speculate that the minor departures from the exact power law scaling for the peak current stem from the null being located at different proximities to the nearest grid point for different values of $N$.

\section{Discussion and conclusions}\label{discuss}
The results presented above clearly demonstrate that the formation of current sheets in response to the compression (or expansion) of a potential magnetic field is by no means a certainty. When the  field contains one or more null points and associated separatrix surfaces, then current sheets are likely to form. This is because the lowest energy state for the compressed (expanded) field will in general have different quantities of magnetic flux within different `topological domains' (whose boundaries are the separatrices), while during the ideal evolution a transfer of flux between these domains is prohibited. There are no such restrictions in the absence of nulls, and our results suggest that in general a force-free field with a finite but spatially-distributed current is accessible under an ideal relaxation. Indeed, \cite{bineau1972} has proven that in the absence of null points, force-free fields do exist in the vicinity of the potential field, i.e.~for sufficiently small $\alpha$ (though the proof does not provide any numerical estimate for how small $\alpha$ should be). 
We note that the finite-amplitude compression of a magnetic field in the manner described herein is rather different to the original mechanism of spontaneous current sheet formation put forward by Parker to explain solar coronal heating. In the study of JL09, the choice of a compression of the domain was simply a method to generate a domain and magnetic topology that are inconsistent with the existence of a smooth equilibrium. However, we have shown that in the absence of null points, no such inconsistency arises.

The reason for the breakdown of the argument made by JL09 has been discussed in detail by \cite{aly2010}. In particular, the assumption made by JL09, based on intuitive arguments, that a topologically untwisted magnetic field must be potential, is shown to be incorrect. We have demonstrated that the resolution of the change of connectivity seen by JL09 is not that space-filling current sheets form in the relaxed state, but that a large spatially-diffuse current develops. We have in addition applied our relaxation technique to the magnetic field presented by \cite{huang2009} -- which also contains no null points. Again, large-scale diffuse current concentrations were observed in the relaxed state, confirming their analysis.
We note that the present implementation of the relaxation negates the objections regarding this counter-example put forward by \cite{janse2010}, who asserted that the field did not conform to the geometry considered in their theory based on the fact that it had infinite extent in the invariant direction. However, here we enclosed the field in a finite box (line-tied on all boundaries, such that during the relaxation the field becomes fully three-dimensional), and have obtained a smooth solution for the relaxed magnetic field that shows good agreement with the results of \cite{huang2009}.

The conditions under which singular current sheets form during the ideal relaxation of a magnetic field remain unknown. What we have demonstrated here is that, in general, the simple compression or expansion of a potential magnetic field without null points does not lead to a lowest energy state containing singular current sheets. 
It is likely that there are certain potential magnetic fields that contain no null points or separatrices but that would develop tangential discontinuities when perturbed, for example by a compression, and then ideally relaxed. It would be a major theoretical breakthrough to understand what additional properties of the magnetic field are required for this to occur.


\acknowledgments

We are grateful to G.~Hornig and A.~L.~Wilmot-Smith for helpful discussions. D.~I.~Pontin gratefully acknowledges the financial support of a Philip Leverhulme Prize. Y.-M. Huang gratefully acknowledges support by the National Science Foundation, Grant No. PHY-0215581 (PFC: Center for Magnetic
Self-Organization in Laboratory and Astrophysical Plasmas)
and the Department of Energy, Grant No. DE-FG02-07ER46372, under the auspice of the Center for Integrated Computation and Analysis of Reconnection and Turbulence.



\appendix

\section{Assessment of errors in the Lagrangian relaxation scheme}\label{appA}
It should be noted that numerical roundoff errors can in principal mean that the value obtained by the numerical method for the level of the $\JJ\times\BB$ force does not always provide an accurate measure of the proximity of the field to force-free equilibrium. The error in the evaluation of $\JJ\times\BB$ tends to build as the grid deformation becomes stronger, as discussed by \cite{pontin2009}.  However, the advantage of the simulations carried out in Section \ref{nonullsec} is that the grid deformation required to reach the equilibrium state turns out to be relatively weak. 
The accuracy of the approximation can be estimated by calculating the current density via a `mimetic' approach. We employ the method described by \cite{pontin2009} to evaluate the Lorentz force as a check on our results. For $N=101$, we find a maximum Lorentz force within $-1\leq x,y\leq 1$ of $1.4\times 10^{-3}$, comparing favourably with the stated value of $1\times10^{-3}$, and indicating that roundoff errors are small.

\end{document}